\documentclass[
    aps,
    prl,
    reprint,
    superscriptaddress,
    preprintnumbers,
    amsmath,
    amssymb,
    floatfix
]{revtex4-2}

\usepackage{graphicx}   
\usepackage{bm} 
\usepackage{dcolumn}
\usepackage{xcolor}   
\usepackage{hyperref}
\usepackage{braket} 
\usepackage{cancel}
\usepackage[normalem]{ulem}
\hypersetup{hidelinks}

\begin{document} 

\newcommand{\nn}{\nonumber\\}

\title{Collider Detector Observables from Lattice Spin Systems}
 
\author{Jo\~{a}o Barata}
\affiliation{CERN, Theoretical Physics Department, CH-1211, Geneva 23, Switzerland}

\author{Ying-Ying Li}
\affiliation{Institute of High Energy Physics, Chinese Academy of Sciences, Beijing 100049, China}

\author{Bo Wang}
\affiliation{Department of Physics, Yale University, New Haven, CT 06511}

\author{Hua Xing Zhu}
\affiliation{School of Physics, Peking University, Beijing 100871, China}
\affiliation{Center for High Energy Physics, Peking University, Beijing 100871, China}

\preprint{CERN-TH-2026-216}

\begin{abstract}
Detector observables pose a fundamental challenge for nonperturbative lattice methods due to their intrinsically real-time and asymptotic nature. We propose quantum simulation platforms as an ideal setting for studying these observables. Using quantum spin systems as a demonstration, we construct lattice detector operators and establish their connection to continuum energy-flow observables. We demonstrate this framework in the $1+1$-dimensional quantum Ising model, where finite lattice calculations show quantitative agreement with conformal field theory expectations at criticality. These results support the interpretation of the lattice flow operators as regulated counterparts of continuum detectors in the scaling and asymptotic limits, providing a concrete route toward realizing high-energy collider physics on table-top quantum platforms. 
\end{abstract}

\maketitle

\textit{Introduction:} One key challenge in studying quantum field theory (QFT) through high-energy collider experiments lies in the fact that only final asymptotic particle states can be directly measured, and the rich dynamics of the underlying theory must be inferred from them. Although the theoretical formulation of high-energy scattering in QFT has been established and extensively explored over the past several decades~\cite{Lehmann:1954rq,Weinberg:1995mt}, a rigorous theoretical description of collider detectors has emerged only more recently~\cite{Hofman:2008ar, Kravchuk:2018htv,Kologlu:2019mfz,Caron-Huot:2022eqs} in a conformal setup. These detectors can be realized as light transforms~\cite{Kravchuk:2018htv} of conserved currents, giving rise to the notion of asymptotic flows of conserved charges. The most extensively studied detector is the \textit{energy-flow} operator \cite{Hofman:2008ar,Sveshnikov:1995vi,Korchemsky:1999kt,Bauer:2008dt}:
\begin{align}\label{eq:def_E_detector}
\mathcal{E}(\hat{n}) \equiv \lim_{r\to\infty} r^{d-1}
\int_0^\infty dt\, \hat{n}^{\,i} T_{0i}(t,r\hat{n}) \, ,
\end{align}
where $d$ corresponds to the number of spatial dimensions, and $\hat n$ is a normal vector to a $d-1$-sphere of radius $r$. The $\mathcal{E}$ operator integrates the energy momentum tensor along light-rays, effectively measuring the energy flow passing through idealized asymptotic detectors. Correlators of such operators~\cite{Basham1978,Basham1979}
\begin{align}\label{eq:correlators}
\left\langle  \mathcal{E}(\hat{n}_1)\cdots\mathcal{E}(\hat{n}_k) \right\rangle
=\frac{
\bra{\Omega} O^\dagger \mathcal{E}(\hat{n}_1)\cdots\mathcal{E}(\hat{n}_k) O\ket{\Omega}}{\braket{\Omega|O^\dagger O|\Omega}}\, ,
\end{align}
evaluated in non-equilibrium states $O|\Omega\rangle$ created by local operator insertions, provide a field-theoretic description of collider measurements, see~\cite{Moult:2025nhu} for a recent review. They characterize how the energy generated by the source is distributed among asymptotic detectors. Multipoint $\mathcal{E}$ correlators thus provide access to the dynamical properties of the underlying QFT through observables with a direct experimental realization~\cite{CMS:2024mlf,ATLAS:2023tgo}.

For non-conformal theories, such as QCD, determining these detector correlators is challenging. Perturbative QCD provides a systematic description in the high-energy regime~\cite{DelDuca:2016csb,Dixon:2018qgp,Dixon:2019uzg,Moult:2018jzp,Ebert:2020sfi,Li:2021zcf,Electron-PositronAlliance:2025fhk}, while their nonperturbative structure has been explored using a range of complementary methods, see e.g.~\cite{Chen:2026hmd,Chang:2025kgq,Chen:2026lsc, Barata:2026pgh, Schindler:2023cww, Belitsky:2001ij,Dokshitzer:1999sh, Riembau:2024tom,Barata:2026eth,Belin:2026wkc}.
First-principles computations of correlation functions involving $\mathcal{E}$ operators would naturally call for a lattice formulation. A conventional Euclidean lattice QFT formulation, however, is not directly applicable due to the intrinsically real-time nature of these operators. In this context quantum simulation~\cite{Banuls:2019bmf,Klco:2021lap,Martinez:2016yna,Bauer:2022hpo,DiMeglio:2023nsa, Fang:2024ple} offers a natural route to directly realize Eq.~\eqref{eq:correlators} in quantum many-body systems with controlled real-time dynamics. While early approaches in this direction have been explored~\cite{Barata:2024apg,Barata:2025jhd,Cao:2026nyj}, a quantitatively systematic treatment of detector correlators in quantum mechanical platforms is lacking.

In this \textit{Letter}, we develop a framework for realizing detector correlators in quantum spin systems, constructing latticized energy flow detectors which can be connected to their continuum formulation in Eq.~\eqref{eq:def_E_detector}. We demonstrate this construction in the $1+1$-dimensional transverse-field Ising model,
\begin{align}
H(g)=-\sum_j Z_jZ_{j+1}-g\sum_jX_j \, ,
\label{eq:ising-hamiltonian}
\end{align}
where $Z_j$/$X_j$ are Pauli operators acting on a chain of $N$ spin-$1/2$ degrees of freedom. At the quantum critical point, $g=1$, the model is described by the $2$d Ising CFT~\cite{Belavin:1984vu,DiFrancesco:1997nk}, and the correlation functions in Eq.~\eqref{eq:correlators} are highly constrained by conformal symmetry, being directly determined by local CFT data~\cite{Hofman:2008ar,Kravchuk:2018htv,Kologlu:2019mfz}. The conformal point thus provides a controlled setting in which lattice detector constructions can be quantitatively benchmarked against the continuum operators. In particular, we show that a dynamical detector formulation, following from microscopic energy conservation, matches the results obtained from static detector constructions, determined by conformal symmetry. The dynamical construction remains well defined away from criticality, thus providing a general lattice realization of detector operators, which can be extended to higher-dimensions.

\textit{2d critical Ising model:} 
We begin by validating the extraction of CFT data from the critical lattice model.
In the continuum theory, Euclidean spacetime can be parametrized by the complex coordinates $z=x+iv\tau$ and $\bar z=x-iv\tau$, where $x$ is the spatial coordinate, $\tau$ is the Euclidean time, and $v$ is the emergent light velocity. Conformal transformations then separate into holomorphic and antiholomorphic sectors whose generators form two commuting Virasoro algebras, see End Matter (EM). After analytic continuation to real-time, these sectors describe right ($\mathcal{R}$) and left ($\mathcal{L}$) moving excitations, respectively.

A local primary field operator $\phi$ in the 2d CFT is specified by conformal weights $(h,\bar h)$, or equivalently by its scaling dimension $\Delta=h+\bar h$ and Lorentz spin $s=h-\bar h$. The Ising CFT has only three primary families: the identity $\mathbf 1$ [$(0,0)_{\mathbf 1}$], the spin field $\sigma$ [$(1/16,1/16)_{\sigma}$], and the energy field $\epsilon$ [$(1/2,1/2)_{\epsilon}$]. 
Acting with the negative modes of the Virasoro generators $L_{-n}$ and $\bar L_{-n}$ ($n>0$) allows one to build families of descendants from the primary operator $\phi$; a descendant with levels $(M,\bar M)$ then has $\Delta=\Delta_\phi+M+\bar M$ and $s=s_\phi+M-\bar M$. To compare this operator content with a finite periodic spin chain, one can quantize the theory on a cylinder of circumference $L=Na_l$, with $z=\exp(2\pi w/L)$, where $w=v\tau+ix$, $x\sim x+L$, and we set the lattice spacing to \(a_l=1\). The finite-size spectrum directly resolves the continuum quantum numbers 
\begin{align}
E-E_0=\frac{2\pi v}{L}\Delta, 
\qquad
P=\frac{2\pi}{L}s \, ,
\label{eq:cft-cylinder-spectrum}
\end{align}
where $E$ and $P$ are the energy and momentum of the corresponding state, and $E_0$ is the ground state energy.

\begin{figure}[t]
  \centering
     \includegraphics[width=\columnwidth]{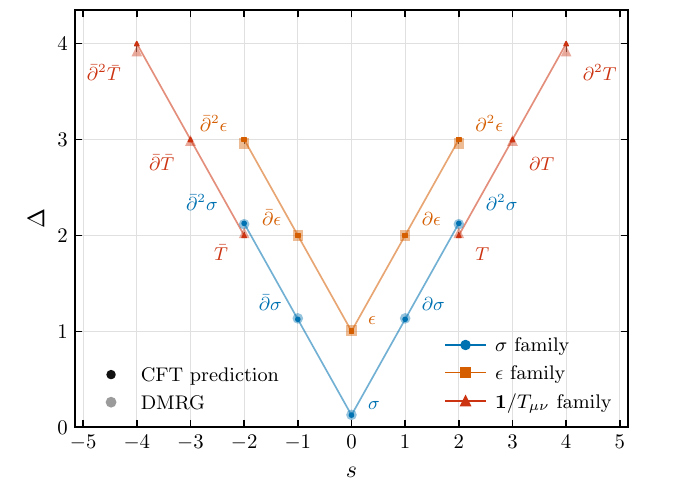}
    \includegraphics[width=\columnwidth]{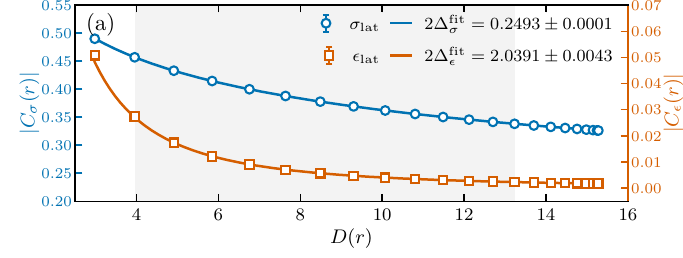}
  \includegraphics[width=\columnwidth]{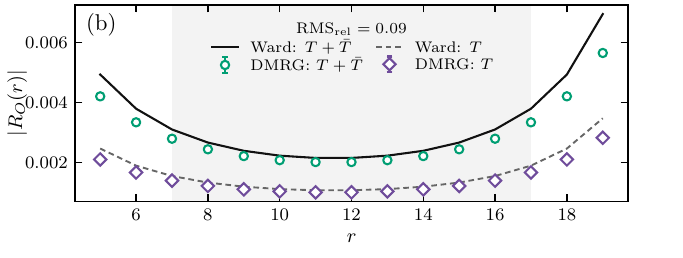}
  \caption{\textbf{Top:} $(\Delta,s)$ decomposition of the CFT operator families. Small solid markers show the CFT predictions; larger translucent markers show the DMRG extraction. \textbf{Bottom:} $(a)$ Spin and energy two-point functions with chord distance fits. Standard error regression uncertainties at fixed \(N\), bond dimension, and fitting window, are quoted. $(b)$ Lattice test of the stress tensor Ward identity.}
  \label{fig:lattice-validation}
\end{figure}

In Fig.~\ref{fig:lattice-validation} (top) we show the $(\Delta,s)$ operator decomposition for the CFT up to second-order descendants of $\sigma$, $\epsilon$, and the stress tensor $T$, which belongs to the identity family, see also~\cite{Zou:2017zce,Milsted:2017csn,Zou:2019dnc,Radicevic:2019mle}. Solid markers denote the CFT levels obtained from the operator-state correspondence in the continuum theory. The lattice values for the scaling dimensions are obtained on a periodic $N=24$ chain at criticality. We determine the ground state $\ket{\Omega}$ and construct, for each lattice operator $O_j$ and momentum $s$, the source operator
$O_s=N^{-1/2}\sum_j e^{2\pi i s(j-1)/N}O_j$. After removing the vacuum component, $O_s|\Omega\rangle\to(1-|\Omega\rangle\langle\Omega|)O_s|\Omega\rangle$, we construct the Krylov space $\mathcal K_O=\mathrm{span}\{O_s|\Omega\rangle,(H-E_0)O_s|\Omega\rangle,\ldots\}$.
We choose Krylov dimensions $K=4$ for the $\sigma$ and $T$ families, and $K=8$ for the $\epsilon$ family,  diagonalizing $H$ projected onto $\mathcal K_O$ to obtain Ritz states with energies $E_n$\footnote{For the $\epsilon$ scalar, the extracted dimensions are
$\Delta_\epsilon^{\rm lat}=1.2686$, $1.0533$, and $1.0095$ for $K=4$, $6$, and $8$, respectively, motivating the choice $K=8$.}. Using Eq.~\eqref{eq:cft-cylinder-spectrum}, we convert these to lattice scaling dimensions and retain the level closest to the corresponding CFT operator, shown as larger translucent markers in Fig.~\ref{fig:lattice-validation} (top). Good agreement with the corresponding continuum CFT predictions can be observed. The lattice operators $O_j$ used for $\sigma$, $\epsilon$ and $T$ are defined below, while their continuum matching and descendant construction are detailed in
the EM. All calculations are performed using matrix product states~\cite{Verstraete:2008cex,Cirac:2020obd} implemented with the \texttt{ITensor} package~\cite{Fishman2022ITensor}. Ground states are obtained using the DMRG algorithm~\cite{White:1992zz,White:1993zza}. For the periodic $N=24$ Ritz calculations, the ground state was obtained with maximum bond dimension $520$, while the Krylov-vector applications used $220$.

At criticality, the ground state correlations of local operators are highly constrained by symmetry. For a scalar primary operator $\phi$, conformal covariance fixes the equal-time cylinder two-point function, up to an operator-dependent normalization, and gives direct access to the scaling dimension $\Delta_\phi$:
\begin{align}
C_\phi(r) \equiv \langle\phi(x_1)\phi(x_2)\rangle_{\rm cyl}
=\frac{B_\phi}{D(r)^{2\Delta_\phi}},
\label{eq}
\end{align}
where $r=|x_1-x_2|$ is the spatial separation between the two operators and
$D(r)=\frac{L}{\pi}\sin\left(\frac{\pi r}{L}\right)$ is the corresponding chord distance on the cylinder. Correlators involving the stress tensor obey the more restrictive conformal Ward identity,
\begin{align}
\left\langle T(z)\prod_i\phi_i(z_i,\bar z_i)\right\rangle
&=\sum_i\left[
\frac{h_i}{(z-z_i)^2}+\frac{1}{z-z_i}\partial_{z_i}
\right]\notag\\
&\quad\times
\left\langle\prod_i\phi_i(z_i,\bar z_i)\right\rangle,
\label{eq:continuum-ward-identity}
\end{align}
with an analogous relation for $\bar T$. This identity ties the
stress tensor three-point function to derivatives of the primary two-point
function and therefore tests not only the scaling exponent but also the lattice
normalization and tensor structure. In
Fig.~\ref{fig:lattice-validation} $(a)$, we evaluate the connected correlators
of the centered lattice representatives
\(\sigma_j^{\rm lat}=Z_j\) and
\(\epsilon_j^{\rm lat}
=X_j-\tfrac12(Z_{j-1}Z_j+Z_jZ_{j+1})\)
in a periodic lattice with $N=48$~\cite{Grinza:2002pi,Caselle:1999bx}. The ground state used for both the correlator and Ward identity calculations was obtained from DMRG using a maximum
bond dimension $1000$.

The correlators are averaged over
12 translated reference positions and fitted over $4\le r\le16$. Panel $(b)$ tests the Ward identity through the connected ratio $R_O(x_2)= \langle\sigma(x_1)O(x_2)\sigma(x_3)\rangle_c/\langle\sigma(x_1)\sigma(x_3)\rangle_c$ for $O = T, \, T+\bar T$. Writing $z_k=\exp(2\pi i x_k/L)$, $z_{ij}=z_i-z_j$, the continuum profiles shown in the figure are
\begin{align}
R_T(x_2)
&=\left(\frac{2\pi}{L}\right)^2
\frac{z_2^2h_\sigma z_{13}^2}{z_{21}^2z_{23}^2}\, ,
\label{eq:ward-profile}
\end{align}
and $R_{T+\bar T}(x_2)=2 {\rm Re} R_T(x_2)$. On the lattice we use the bond-centered energy density $h_j=-Z_jZ_{j+1}-\tfrac{g}{2}(X_j+X_{j+1})$ and continuity current $j_j^E=i[h_j,h_{j+1}]$, to construct the energy momentum tensor (see EM),
\begin{align}
(T+\bar T)_j=\frac{2\pi}{v}h_j,\qquad
T_j=\frac{\pi}{v}h_j+\frac{\pi}{v^2}j_j^E,
\label{eq:ward-lattice-map}
\end{align}
where $v=2$, following from the dispersion relation
$\omega(k)=2\sqrt{1+g^2-2g\cos k}\approx 2|k|$. The gray region marks the bulk comparison
window $7\le r\le17$, where the insertion is kept away from both external
spin operators and short-distance lattice effects are reduced, using $x_1=0$ and $x_3=L/2=24$, and the remaining coordinate is parametrized as $x_2=r+\tfrac12$. Points outside this region are shown but are not included in the quoted relative root mean square deviation $\mathrm{RMS}_{\rm rel}$ between the lattice data and the CFT Ward identity prediction. These results justify the construction of lattice current and energy momentum tensor, and thus enable the direct construction of lattice detector operators and the calculation of their correlations, as we show below.

\textit{Lattice realization of energy flow operators:} The decomposition of the CFT into left- and right-moving sectors permits complementary static and real-time realizations of detector operators. 
Since the celestial sphere reduces to two points in $1+1$-dimensions, at the gapless point the Hamiltonian and total energy current decompose into commuting chiral zero modes, 
\begin{align}
H-E_0&=H_\mathcal{R}+H_\mathcal{L},
\qquad
\frac{J_E}{v}=H_\mathcal{R}-H_\mathcal{L},
\end{align}
where $J_E=\sum_j j_j^E$. The local continuity equation,
$\dot h_j=j_{j-1}^E-j_j^E$, then gives the static lattice detector operators
\begin{align}
\mathcal E_{\mathcal{R}[\mathcal{L}]}^{\rm lat}
&=\frac12\left[(H-E_0)\pm\frac{J_E}{v}\right],
\label{eq:lattice-zero-modes}
\end{align}
with the upper (lower) sign for $\mathcal R$ ($\mathcal L$). For a localized packet $|\psi\rangle$ with excitation energy $E_\psi=\langle H-E_0\rangle_\psi$, the corresponding energy fraction is
\begin{align}
F_\mathcal{R}^{\rm lat}[\psi]
&=\frac{\langle\mathcal E_\mathcal{R}^{\rm lat}\rangle_\psi}{E_\psi}
=\frac12\left[1+\frac{\langle J_E\rangle_\psi}{vE_\psi}\right].
\label{eq:stationary-fraction-lat}
\end{align}

The previous construction relies on the gapless point relation between energy and current, and is therefore valid only at the critical point. Nonetheless, one can construct a lattice representation of Eq.~\eqref{eq:def_E_detector} without knowledge of the conformal theory's properties, only using energy conservation. 
For any cut position $b$ on
the open chain, define the left and right local Hamiltonians
$Q_{\mathcal L}(b)$ and $Q_{\mathcal R}(b)$ by symmetrically splitting the cut term such that $Q_\mathcal{R}(b)+Q_\mathcal{L}(b)=H$.
The continuity equation then gives
$i[H,Q_\mathcal{R}(b)]=j_{b}^E$ and $i[H,Q_\mathcal{L}(b)]=-j_{b}^E$. For detectors placed at distinct cuts $b_a$, the outward current is
$j_a^{\rm out}(b_a,t)=\pm j_{b_a}^E(t)$, with the upper (lower) sign for
$a=\mathcal{R}$ ($\mathcal{L}$). Integrating to the endpoint $t_\star$ gives
the lattice identity corresponding to the continuum detector in Eq.~\eqref{eq:def_E_detector}:
\begin{align}
K_a(b_a,t_\star)
&\equiv U^\dagger(t_\star)Q_a(b_a)U(t_\star)-Q_a(b_a),\notag\\
&=\int_0^{t_\star} dt\,j_a^{\rm out}(b_a,t)\, .
\label{eq:regional-detector}
\end{align}
For a localized packet $|\psi\rangle$, we can thus define the \textit{window} one-point detector observable $F_a^{\rm win}[\psi]
=\frac{\langle K_a(b_a,t_\star)\rangle_\psi}{E_\psi}$. Equation~\eqref{eq:regional-detector} is an exact lattice identity for every value of $g$, as $K_a(b,t)$ measures the net energy transported through the corresponding cut. At the critical point, this operator can be identified
with a regulated realization of the conformal light-ray detector $\mathcal E_a$.

\begin{figure*}[t]
  \centering
  \includegraphics[width=0.49\textwidth]{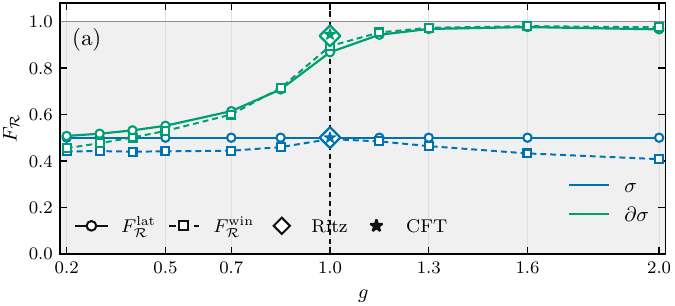}\hfill
  \includegraphics[width=0.49\textwidth]{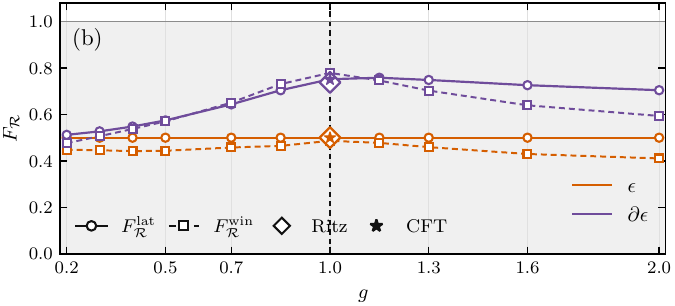}
  \caption{One-point energy-flow across $g$ for (a) the spin and (b) the energy conformal families, obtained from the static construction (circles), window construction (squares), direct measurement on the Ritz states (diamonds) and CFT prediction (stars). 
  }
  \label{fig:one-point-flow}
\end{figure*}

The interpretation of $K_a(b_a,t_\star)$ in Eq.~\eqref{eq:regional-detector} as a detector
at null infinity requires dealing with the scaling and late-time limits. The continuum detector is approached when the scaling hierarchy $a_l\ll\sigma_x\ll R_{\rm det}\ll N/2$ is satisfied, where $\sigma_x$ is the packet width, and $R_{\rm det}$ the source--detector separation. The integration endpoint is chosen after the pulse has crossed the detector at $t_{\rm pass}$ but
before boundary reflections reach it at $t_{\rm reflection}$, i.e.
$t_{\rm pass}<t_\star<t_{\rm reflection}$.
Finite-size calculations therefore retain systematic dependence on $t_\star$ and $\sigma_x$, in addition to errors from the finite time step in the real-time evolution; we do not attempt a continuum or null-infinity extrapolation in the present work. To compare the static and window constructions on equal footing, we scan the coupling $g$ on an $N=24$ open chain using the same
centered packets for the primary fields,
\begin{align}
|\psi_O\rangle
&=\mathcal N_O(1-|\Omega\rangle\langle\Omega|)\notag\\
&\quad\times\sum_j e^{-(j-x_0)^2/(2\sigma_x^2)}
\cos[k_0(j-x_0)]\,O_j|\Omega\rangle,
\label{eq:main-packet-source}
\end{align}
with $\sigma_x=2.5$ and $k_0=0.6$. For the descendants, we instead use a complex phase $e^{ik_0(j-x_0)}$ (rather than the cosine) and apply the centered difference to the full packet weights, as described
in the EM, with $k_0=0.7$. The cosine produces a reflection-even primary packet with equal
$\pm k_0$ components, whereas the complex phase selects one momentum direction for the chiral descendant. For the $\epsilon$ family a short \textit{filtering} step, \(
|\psi_\epsilon(\tau)\rangle\propto
e^{-\tau(H-E_0)}|\psi_\epsilon\rangle ,
\) using \(\tau=0.6\), is introduced to remove admixture with undesired ultraviolet operators. Unlike the conformal-collider preparation used in~\cite{Hofman:2008ar}, our packets have finite spatial extent and bandwidth, with momentum peaks centered near $\pm k_0$; the associated finite bandwidth effects are quantified in the EM.

Figure~\ref{fig:one-point-flow} compares the two constructions for the same packet in Eq. (\ref{eq:main-packet-source}) with the operator
$O$ taken to be the $\sigma$ and $\epsilon$ primaries and their first
right-moving descendants.
Circles denote the static construction $F_{\mathcal R}^{\rm lat}[\psi_O]$, while
squares denote the window construction $F_{\mathcal R}^{\rm win}[\psi_O]$. 
For each coupling, the $N=24$ open-chain ground state was obtained from DMRG with a maximum bond dimension of $96$.
The window detector is placed at $b_{\mathcal
R}=18$ on the $N=24$ open chain and evolved with time step $\delta t=0.08$ using the TEBD algorithm~\cite{Vidal:2003lvx} with maximum bond dimension of $96$; the integration endpoint $t_\star$ is selected after the outgoing pulse crosses the cut but before boundary reflections return, see EM. The gray bands
show the positivity bounds for the continuum operators.

Figure~\ref{fig:one-point-flow} also shows the exact CFT values (stars), and the estimate obtained from the lattice Ritz states (open diamonds) \footnote{For the Ritz estimate we used $N=20$.}. The CFT prediction follows from the state-operator correspondence, which associates each operator \(O\) with a cylinder eigenstate \(\ket{\Psi_O}=\ket{h,\bar h}\), giving
\begin{align}
F^{\rm CFT}_\mathcal{R}[\Psi_O]
&=\frac{h}{h+\bar h}
=\frac{\Delta+s}{2\Delta},
\label{eq:stationary-fraction}
\end{align}
with $F^{\rm CFT}_\mathcal{L}[\Psi_O]=1-F^{\rm CFT}_\mathcal{R}[\Psi_O]$. This relation follows from the integrated stress tensor Ward identity, more concretely $ F^{\rm CFT}_\mathcal{R}[\Psi_\sigma]=F^{\rm CFT}_\mathcal{R}[\Psi_\epsilon]=\frac12$, $F^{\rm CFT}_\mathcal{R}[\Psi_{\partial\sigma}]=\frac{17}{18}$, and $F^{\rm CFT}_\mathcal{R}[\Psi_{\partial\epsilon}]=\frac{3}{4}$.
The Ritz and packet calculations use complementary state preparations: the Ritz vector approximates a cylinder eigenstate, whereas the localized packet superposes several cylinder levels. At criticality, both give results
close to the CFT predictions, up to systematic deviations related to the residual packet dependence arising from its finite bandwidth and detector separation, thus validating our lattice detector construction. 
As the static construction relies
on the critical chiral decomposition, its continuation
away from criticality is shown here for illustration only. By contrast, $F_{\mathcal R}^{\rm win}$ remains an operational detector throughout the coupling scan.

This treatment extends to determine the two-point function of lattice detector operators:
\begin{align}
C_{\mathcal{RL}}^{\rm lat}[\psi]
&=\frac{\langle\{\mathcal E_\mathcal{R}^{\rm lat},
\mathcal E_\mathcal{L}^{\rm lat}\}\rangle_\psi}{2E_\psi^2}
\nn
&=\frac{\langle(H-E_0)^2\rangle_\psi
-\langle J_E^2\rangle_\psi/v^2}{4\langle(H-E_0)\rangle_\psi^2
},
\nn
C_{\mathcal{RL}}^{\rm win}[\psi]
&=\frac{\operatorname{Re}\langle
K_\mathcal{R}(b_\mathcal{R}, t_\star)\psi|K_\mathcal{L}(b_\mathcal{L},t_\star)\psi\rangle}
{E_\psi^2}.
\label{eq:window-two-point}
\end{align} 
As the two-point correlator probes a second moment of the outgoing energy, it is more sensitive than the one-point fraction to the bandwidth of the prepared state and to incomplete finite-time capture. 
In the CFT, since $|\Psi_O\rangle$ is an eigenstate of both
chiral Hamiltonians, the two-point correlator factorizes as
\begin{align}
C_{\mathcal{RL}}^{\rm CFT}[\Psi_O]
=F^{\rm CFT}_\mathcal{R}[\Psi_O]F^{\rm CFT}_\mathcal{L}[\Psi_O]
=\frac{\Delta^2-s^2}{4\Delta^2}.
\label{eq:cft-two-point}
\end{align}
Unitarity gives $h,\bar h\geq0$, or $|s|\leq\Delta$, and hence
$0\leq F_{\mathcal{R},\mathcal{L}}^{\rm CFT}[\Psi_O]\leq1$. Given that $F_\mathcal{R}^{\rm CFT}[\Psi_O]+F_\mathcal{L}^{\rm CFT}[\Psi_O]=1$, $0\leq C_{\mathcal{RL}}^{\rm CFT}[\Psi_O]\leq1/4$. A purely chiral state saturates the lower bound, whereas any scalar state, for which $h=\bar h$, saturates
the upper bound, i.e. $C_{\mathcal{RL}}^{\rm CFT}[\Psi_{\sigma,\epsilon}]=1/4$. For a finite packet, however, the state generally superposes multiple chiral-energy sectors, so the factorization above need not hold, and the two-point correlator is not necessarily bounded by $1/4$.

\begin{figure}[t]
  \centering
  \includegraphics[width=1\columnwidth]{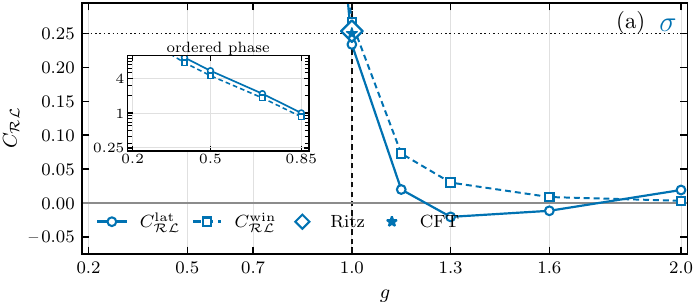}
  \includegraphics[width=1\columnwidth]{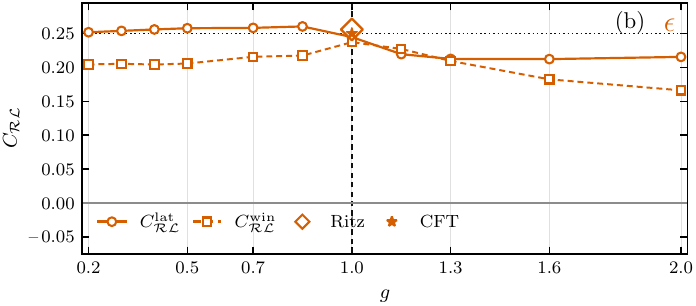}
  \caption{$C_\mathcal{RL}$ correlators for primary packets in (a) the
  spin and (b) the energy family. Notation follows Fig.~\ref{fig:one-point-flow}.
  }
  \label{fig:two-point-flow}
\end{figure}

Figure~\ref{fig:two-point-flow} compares the different constructions of the
two-point correlator, following the conventions, methods, and notation of
Fig.~\ref{fig:one-point-flow}. For all window results, the detector cuts are fixed at $b_{\mathcal L}=5$ and $b_{\mathcal R}=18$.
At criticality, the static and window results are, 
$(C_{\mathcal R\mathcal L}^{\rm lat},
 C_{\mathcal R\mathcal L}^{\rm win})[\psi_\sigma]=(0.2339,0.2661)$ and
$(C_{\mathcal R\mathcal L}^{\rm lat},
 C_{\mathcal R\mathcal L}^{\rm win})[\psi_\epsilon]=(0.2447,0.2372)$, in good agreement with
the CFT prediction, and predictions obtained from Ritz states which are $C_{\mathcal{RL}}[\Psi_\sigma]=0.2535$ and
$C_{\mathcal{RL}}[\Psi_\epsilon]=0.2560$.
To quantify finite-time detector
capture, we define
$R_n=\langle (K_{\mathcal R}+K_{\mathcal L})^n\rangle/\langle (H-E_0)^n \rangle$, which tests the recovery of the 
energy moments. Their values,
$(R_1,R_2)_\sigma=(0.9884,1.0930)$ and $(R_1,R_2)_\epsilon=(0.9737,1.0395)$, explain the larger window deviation in the spin channel. These values use family-dependent endpoints; in the EM we select a common criterion instead. Away from criticality, the $\epsilon$ curves remain smooth, whereas the weak-coupling growth of $\sigma$ curves arises from its overlap with the finite-size parity doublet, which suppresses $E_\psi^2$ and renders the ratio ill-conditioned.

\textit{Conclusion:} We have constructed asymptotic energy flow detector operators in the $1+1$-dimensional Ising model, building a quantitative connection to their QFT counterparts. At criticality, we recovered the CFT predictions for the local correlation functions and the operator decomposition, allowing for a lattice evaluation of detector correlators. While many of the elements discussed here are specific to two-dimensional theories, the dynamical detector operator remains well defined away from criticality, and follows only from microscopic energy conservation. Its interpretation there is that of an energy detector, while the  identification with a conformal light-ray operator applies at the critical scaling limit. Its construction admits an extension to higher dimensions, requiring no prior knowledge of the theory's spectrum.

Energy flow detector observables require simulating an extensive volume for a time of order its linear size, i.e. $V\sim R_{\rm det}^d$ and $t_{\star}\sim R_{\rm det}/v$. This quickly becomes intractable for classical tensor network-based simulations in $2+1$-dimensions, while digital quantum simulations will face the challenge of large system sizes and long real-time evolution, clearly beyond the capabilities of current or near term devices~\cite{Bauer:2022hpo,DiMeglio:2023nsa,Fang:2024ple}. 
One interesting and natural alternative route is analog quantum simulation. Recent experiments have studied 
the spectrum of the one-dimensional Ising chain in Rydberg
arrays, demonstrating the state--operator correspondence of two-dimensional CFTs~\cite{Sun:2026cft}, and there has been related recent progress in the study of the three dimensional Ising model~\cite{Ebadi:2020ldi,Scholl:2020hzx,Manovitz:2024hif,Fang:2024uyf}. Realizing our detector construction in such experiments would provide a direct test of CFT beyond scaling dimensions. Moreover, using the dynamical detector operator construction, this would provide a route to study detector operators in non-conformal theories, allowing the realization of collider physics on table-top quantum platforms.

\textbf{Acknowledgments:} We thank L. Tizzano for pointing out~\cite{Grinza:2002pi} to us and for useful discussions.
YYL is supported by the National Key R\&D Program of China (Grant Nos. 2025YFA1614200), the National Science Foundation of China under Grant Nos. 12522509. YYL would also like to thank the Aspen Center for Physics, which is supported by National Science Foundation grant No. PHY-2210452, where part of this work has been done. BW is partially supported by the DOE Early Career Award DE-SC0025581, the Sloan Foundation, and the Simons Collaboration on Confinement and QCD Strings. HXZ is supported by
the National Natural Science Foundation of China under grant No. 12425505. 

\bibliographystyle{apsrev4-2}
\bibliography{reference_EEC.bib}

\section{End Matter}

\subsection*{Primer on the 2d Ising CFT}

The local conformal symmetry is generated by the Laurent modes of the
holomorphic and antiholomorphic stress tensors,
\begin{align}
T(z)=\sum_{n\in\mathbb Z}L_n z^{-n-2},
\qquad
\bar T(\bar z)=\sum_{n\in\mathbb Z}\bar L_n\bar z^{-n-2}.
\end{align}
These modes form two commuting Virasoro algebras,
\begin{align}
[L_m,L_n]
&=(m-n)L_{m+n}
 +\frac{c}{12}m(m^2-1)\delta_{m+n,0},
\notag\\
[\bar L_m,\bar L_n]
&=(m-n)\bar L_{m+n}
 +\frac{c}{12}m(m^2-1)\delta_{m+n,0},
\notag\\
[L_m,\bar L_n]&=0,
\label{eq:em-virasoro}
\end{align}
with central charge $c=1/2$. A primary state
$|\phi\rangle$ satisfies
\begin{align}
&L_0|\phi\rangle=h|\phi\rangle,\quad
\bar L_0|\phi\rangle=\bar h|\phi\rangle,\\ \notag
&L_{n>0}|\phi\rangle=\bar L_{n>0}|\phi\rangle=0.
\end{align}
Descendants are obtained by acting with the lowering operators
$L_{-n}$ and $\bar L_{-n}$. A descendant at levels
$(\ell,\bar\ell)$ has $\Delta=h+\bar h+\ell+\bar\ell$ and $s=h-\bar h+\ell-\bar\ell$. In the identity family, the first nontrivial descendants are
\begin{align}
T=L_{-2}\mathbf 1,\qquad
\bar T=\bar L_{-2}\mathbf 1,
\end{align}
with $(\Delta,s)=(2,2)$ and $(2,-2)$, respectively. Their first
chiral descendants are $\partial T$ and $\bar\partial\bar T$, with
$(\Delta,s)=(3,3)$ and $(3,-3)$. For a scalar primary,
$L_{-1}^{n}\phi$ and $\bar L_{-1}^{n}\phi$ correspond to
$\partial^n\phi$ and $\bar\partial^n\phi$.

At the critical point, a local lattice operator
$O_j^{\rm lat}$ admits an infrared expansion in continuum CFT operators,
\begin{align}
O_j^{\rm lat}
=
\sum_{\alpha}
A_{\alpha}\,
a_{l}^{\Delta_\alpha}
\mathcal O_\alpha(x_j),
\qquad x_j=j\,a_{l},
\label{eq:em-generic-lattice-matching}
\end{align}
where the sum runs over CFT operators compatible with the microscopic symmetries and lattice quantum numbers of $O_j^{\rm lat}$. The scaling dimensions $\Delta_\alpha$ determine the powers of the lattice spacing, while the coefficients $A_\alpha$ are nonuniversal. At long distances, the expansion is dominated by the operator of the lowest scaling dimension allowed in that channel. For the lattice sources used in this work, the leading matching relations are
\begin{align}
Z_j
&=
A_\sigma a_{l}^{1/8}\sigma(x_j)+\cdots,
\notag\\
\epsilon_j^{\rm lat}
&\equiv
X_j-\frac12\left(Z_{j-1}Z_j+Z_jZ_{j+1}\right)
\notag\\
&=
A_\epsilon a_{l}\epsilon(x_j)+\cdots ,
\label{eq:em-primary-dictionary}
\end{align}
where the ellipses denote higher-dimensional operators allowed by the same symmetries. The first relation follows because $Z_j$ is odd under the global $\mathbb Z_2$ symmetry and therefore couples at leading order to the spin primary $\sigma$. The centered even combination $\epsilon_j^{\rm lat}$ is chosen to suppress its identity component and couples at leading order to the energy primary $\epsilon$. The nonuniversal amplitudes $A_\sigma$ and $A_\epsilon$ cancel from the normalized observables considered in the main text. The stress tensor is matched using the centered bond-energy density $h_j$ (after subtracting the local expectation value) and the lattice energy current $j_j^E$ (see next section):
\begin{align}
T_j
=\frac{\pi}{v} h_j+\frac{\pi}{v^2}j_j^E\, , \quad
\bar T_j
=\frac{\pi}{v}h_j-\frac{\pi}{v^2}j_j^E.
\label{eq:em-stress-dictionary}
\end{align}
Equivalently, the first spatial descendants of $\sigma$ and $\epsilon$ are represented by centered lattice differences:
\begin{align}
\delta_x O_j^{\rm lat}
\equiv
\frac{O_{j+1}^{\rm lat}-O_{j-1}^{\rm lat}}{2},
\end{align}
for which the primary matching relations imply
\begin{align}
\delta_x Z_j
=
A_\sigma a_l^{9/8}\,
\partial_x\sigma(x_j) \, , \quad \delta_x\epsilon_j^{\rm lat}
=
A_\epsilon a_l^{2}\,
\partial_x\epsilon(x_j) \, ,
\label{eq:em-descendant-dictionary}
\end{align}
where $\partial_x\mathcal O=\partial\mathcal O+
\bar\partial\mathcal O$. The two chiral descendants are separated by
momentum projection: the sectors $s=+1$ and $s=-1$ select
$\partial\mathcal O$ and $\bar\partial\mathcal O$, respectively, up to
the convention on the lattice momentum sign. Thus,
\begin{align}
\delta_x O_{s=+1}^{\rm lat}
&\longleftrightarrow
A_O a_l^{\Delta_O+1}\partial O \,, 
\\ \notag
\delta_x O_{s=-1}^{\rm lat}
&\longleftrightarrow
A_O a_l^{\Delta_O+1}\bar\partial O .
\end{align}
For the localized sources, the centered difference is applied to the packet weights rather than to the microscopic operator:
\begin{align}
\sum_j f_j\,\delta_x O_j^{\rm lat}
=
-\sum_j(\delta_x f_j)\,O_j^{\rm lat}\, .
\end{align}
The two prescriptions therefore prepare the same descendant packet, assuming that boundary effects coming from the discrete integration by parts are not important.

\subsection*{Microscopic continuity equation}

The real-time detector follows from an exact finite-lattice identity.  We use the symmetric bond density
\begin{align}
h_j=-Z_jZ_{j+1}-\frac{g}{2}(X_j+X_{j+1}),
\qquad H=\sum_jh_j ,
\label{eq:em-bond-density}
\end{align}
for which the Heisenberg equation is
\begin{align}
\dot h_j&=i[H,h_j]=j_{j-1}^E-j_j^E,\notag\\
j_j^E&=i[h_j,h_{j+1}]
=g(Y_{j+1}Z_{j+2}-Z_jY_{j+1}).
\label{eq:em-lattice-current}
\end{align}
For an open chain, the half fields missing from the end bonds are restored by
\begin{align*}
H_{\rm open}
&=-\frac{g}{2}(X_1+X_N)+\sum_{j=1}^{N-1}h_j,\notag\\
Q_\mathcal{L}(b)&=-\frac{g}{2}X_1+\sum_{j=1}^{b}h_j,\quad
Q_\mathcal{R}(b)=\sum_{j=b+1}^{N-1}h_j-\frac{g}{2}X_N .
\label{eq:em-regional-hamiltonians}
\end{align*}
These definitions obey $Q_\mathcal{L}(b)+Q_\mathcal{R}(b)=H_{\rm open}$ and
\begin{align}
i[H,Q_\mathcal{R}(b)]=j_b^E ,\qquad i[H,Q_\mathcal{L}(b)]=-j_b^E.
\end{align}
Integrating these relations gives
Eq.~\eqref{eq:regional-detector}. A local improvement
$h_j\to h_j+b_{j+1}-b_j$ changes pointwise currents by a lattice derivative,
but leaves the integrated transported energy invariant, when the boundary terms vanish.  

For completeness, the $N=48$ correlator and Ward identity calculations use 20 DMRG sweeps, and maximum bond dimension $1000$, while the $N=24$ open-chain calculations use maximum bond dimension $96$. Real-time evolution
uses second-order TEBD with $\delta t=0.08$, maximum bond dimension $96$, and cutoff $10^{-9}$. 

\begin{figure}[h]
  \centering
   \includegraphics[width=.75\columnwidth]
  {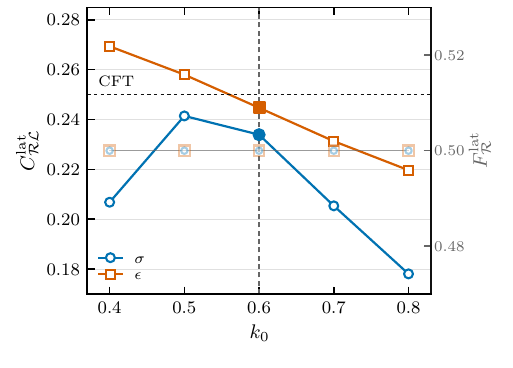}
  \includegraphics[width=.75\columnwidth]
  {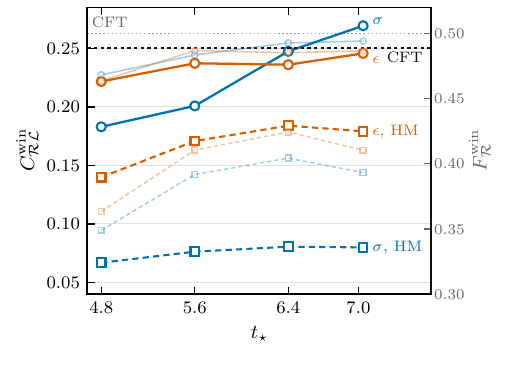}
  \caption{Comparison of packet preparations at $g=1$.
(a) Static two-point correlator $C_{\mathcal R\mathcal L}^{\rm lat}$ (left axis) and one-point fraction $F_{\mathcal R}^{\rm lat}$ (right axis) as functions of the carrier momentum $k_0$. Open markers denote the $\sigma$ and $\epsilon$ packets; filled markers and the vertical line identify the choice $k_0=0.6$ used in the main text. (b) Real-time window correlator $C_{\mathcal R\mathcal L}^{\rm win}$ (left axis) and corresponding one-point fraction $F_{\mathcal R}^{\rm win}$ (right axis) for the spatially localized packet (solid curves and circles) and the Hofman--Maldacena (HM) packet (dashed curves and squares). The one-point data are shown in lighter shades. All results use an $N=24$ open chain, $\sigma_x=2.5$, detector cuts $(b_{\mathcal L},b_{\mathcal R})=(5,18)$, and time step $\delta t=0.08$. The horizontal reference lines denote the scalar CFT values $C_{\mathcal R\mathcal L}=1/4$ and
$F_{\mathcal R}=1/2$.}
  \label{fig:em-packet-comparison}
\end{figure}

\subsection*{Comparison of packet preparations}

The main text uses the spatially localized packet for primaries
\begin{align}
|\psi_O\rangle
&\propto
(1-|\Omega\rangle\langle\Omega|)
\sum_j e^{-(j-x_0)^2/(2\sigma_x^2)}
\notag\\
&\quad\times
\cos[k_0(j-x_0)]\,O_j|\Omega\rangle ,
\label{eq:em-main-packet}
\end{align}
with $\sigma_x=2.5$. The momentum $k_0$ controls both the propagation velocity and the energy and current moments entering the two-detector correlator. Figure~\ref{fig:em-packet-comparison}(a) shows the dependence of $C_{\mathcal{RL}}^{\rm lat}$ on $k_0$, with all other parameters held fixed.
Filled markers and the vertical line identify the main-text choice $k_0=0.6$. Although $k_0\simeq0.5$ gives static values closest to $1/4$, $k_0=0.6$ is the smallest carrier for which both families satisfy $R_1\geq0.95$ and $|R_2-1|\leq0.07$ at the common endpoint
$t_\star=6$. The main text scan uses
independently selected values. The residual $k_0$ dependence is a finite-bandwidth and finite-volume systematic: different carriers
superpose different energy and current sectors. This systematic error should decrease as
$N$, $\sigma_x$, and the source--detector separation are increased
together, producing a narrower packet with a longer evolution window. A
controlled extrapolation of these effects requires a systematic finite-size,
packet-width, and detector-window study, which lies beyond the scope of the
present exploratory work.

The packet in Eq.~\eqref{eq:em-main-packet} has controlled spatial and
momentum support, but it is not narrow in energy. A closer lattice analogue of the Hofman--Maldacena (HM) preparation in~\cite{Hofman:2008ar} starts from a zero-momentum spatial Gaussian and applies an energy-selective temporal
filter:
\begin{align}
|\psi_{O,q}^{\rm HM}\rangle
&\propto
e^{-\sigma_t^2(H-E_0-q)^2/4}
(1-|\Omega\rangle\langle\Omega|)
\notag\\
&\quad\times
\sum_j e^{-(j-x_0)^2/(2\sigma_x^2)}
O_j|\Omega\rangle .
\label{eq:em-hm-packet}
\end{align}
Here $q$ is the target excitation energy and
$\delta E\sim\sigma_t^{-1}$ and the maximum bond dimension is $128$.

Figure~\ref{fig:em-packet-comparison}(b) compares the HM packet, with $(q,\sigma_t)=(1.6,3)$ (dashed curves and squares), with the cosine packet used in the main text (solid curves and circles), as a function of $t_\star$. Each curve is obtained from an independent time evolution using the same chain and detector parameters. The horizontal line denotes the scalar CFT value $C_{\mathcal{RL}}=1/4$. Despite its narrower energy distribution, the HM packet gives poorer detector capture on this finite chain. Its temporal extent $v\sigma_t$ is comparable to the available source--detector separation, so the narrow-energy and asymptotic-detector limits can not be realized simultaneously at $N=24$. The present results consequently retain some dependence on the packet preparation.

\end{document}